\RequirePackage{lineno}
\documentclass[twocolumn,letterpaper,aps,prc,longbibliography,superscriptaddress,nofootinbib,floatfix]{revtex4-1}

\usepackage{graphicx}	
\usepackage{appendix}
\usepackage{float}
\usepackage{bbm}
\usepackage{hyperref}
\usepackage{amsmath}
\usepackage{verbatim}
\usepackage{xspace}

\hypersetup{
    colorlinks=true,
    linkcolor=blue,
    filecolor=blue,
    urlcolor=blue,
    citecolor=blue,
}

\usepackage{xspace}	

\newcommand{\pt}{\mbox{$p_T$}\xspace}

\newcommand{\pbpb}{\mbox{Pb$+$Pb}\xspace}

\newcommand{\auau}{\mbox{Au$+$Au}\xspace}

\newcommand{\sonic}{\mbox{{\sc sonic}}\xspace}

\newcommand{\ipglasma}{\mbox{{\sc ip-glasma}}\xspace}
\newcommand{\ipjazma}{\mbox{{\sc ip-jazma}}\xspace}

\newcommand{\magma}{\mbox{{\sc magma}}\xspace}
\newcommand{\trento}{\mbox{{\sc trento}}\xspace}

\begin{document}

\title{Gluonic Hot Spot Initial Conditions in Heavy-Ion Collisions}

\newcommand{\colorado}{University of Colorado, Boulder, Colorado 80309, USA}
\affiliation{\colorado}

\newcommand{\pusan}{Pusan National University, Busan, 46241, South Korea}

\author{R.~Snyder} \affiliation{\colorado}
\author{M.~Byres} \affiliation{\colorado}
\author{S.H.~Lim} \affiliation{\pusan}
\author{J.L.~Nagle} \affiliation{\colorado}

\date{\today}

\begin{abstract}
The initial conditions in heavy-ion collisions are calculated in many different frameworks.    The importance of nucleon position fluctuations within the nucleus and sub-nucleon structure has been established when modeling initial conditions for input to hydrodynamic calculations.   However, there remain outstanding puzzles regarding these initial conditions, including the measurement of the near equivalence of the elliptical $v_{2}$ and triangular $v_{3}$ flow coefficients in ultra-central 0-1\% \pbpb collisions at the LHC.   Recently a calculation termed \magma incorporating gluonic hot spots via two-point correlators in the Color Glass Condensate framework, and no nucleons, provided a simultaneous match to these flow coefficients measured by the ATLAS experiment, including in ultra-central 0-1\% collisions.    
Our calculations reveal that the \magma initial conditions do not describe the experimental data when run through full hydrodynamic \sonic simulations or when the hot spots from one nucleus resolve hot spots from the other nucleus, as predicted in the Color 
Glass Condensate framework.   We also explore alternative initial condition calculations and discuss their implications.
\end{abstract}


\maketitle



\section{Introduction}
\label{sec:intro}

The physics underlying the first fraction of a fm/$c$ in heavy-ion collisions is of fundamental interest in its own right, while also a necessary input in order to extract properties of the created quark-gluon plasma (QGP) that evolves from this initial state~\cite{Heinz:2013th,Romatschke:2017ejr}.    There are innumerable modelings of the said initial state ranging from claimed {\it{ab initio}} calculations to phenomenological parameterizations~\cite{ALBACETE20141,Lappi:2009fq,Miller:2007ri}.   A major advance in the field more than a decade ago was the incorporation of nucleon position fluctuations via Monte Carlo Glauber calculations~\cite{Miller:2007ri} and subsequently the realization that odd flow coefficients would be non-zero~\cite{Alver:2010gr}.   Most recently it has become clear that sub-nucleon structure is necessary to understand data in proton-proton and proton-nucleus collisions~\cite{Mantysaari:2017cni,Romatschke:2017ejr}, as well as collisions of deformed nuclei such as Uranium-Uranium~\cite{Adamczyk:2015obl}.   

Monte Carlo Glauber code including nucleon and constituent quarks is now publicly available~\cite{Loizides:2016djv}.   
Such calculations have been incorporated into multiple frameworks, including the often used 
\trento~\cite{Moreland:2014oya} model.    In this framework, each incoming nucleon or sub-nucleon is modeled via a two-dimensional Gaussian distribution and the deposited energy is proportional to the square root of the local projectile density times the local target density (in the \trento $p=0$ mode).    Another such framework is \ipjazma~\cite{Nagle:2018ybc}, where the deposited energy can be chosen as the product of local projectile density times the local target density or the square root, as in the \trento model.   Both examples are purely phenomenological; in fact, the \trento model has been used in Bayesian analyses in an attempt to constrain the initial state parameters~\cite{Bernhard:2019bmu}.   A recent comparison of these scalings for multiplicity distributions is detailed in Ref.~\cite{Carzon:2020ohg}.

In contrast, in the weakly-coupled limit, one can in principle calculate the initial conditions in the so-called Color Glass Condensate (CGC) framework (also referred to as the saturation framework) -- for useful reviews see Refs.~\cite{Gelis:2012ri,doi:10.1146/annurev.nucl.010909.083629,Venugopalan_2006}.  Although the calculation is termed {\it ab initio}, it is an effective theory in the limit as the coupling $\alpha_{s}$ goes to zero and for high gluon occupation number, and thus its applicability in the heavy-ion collision regime at RHIC and the LHC is unclear.     Regardless, within this framework one can assume the projectile and target nuclear color charge densities are described by a local saturation scale $Q_{s}$ and then the deposited energy is proportional to the product of the projectile and target color charge densities~\cite{Lappi:2006hq,Romatschke:2017ejr}.   It is notable that as derived in Ref.~\cite{Romatschke:2017ejr}, this simple product is also the result for the deposited energy in the strongly-coupled limit.
The \ipglasma code~\cite{Schenke:2012wb} provides a Monte Carlo framework for the calculation of initial conditions via this CGC effective theory.   The calculation starts with Monte Carlo Glauber with nucleons or sub-nucleons and then associates a local saturation scale with a two-dimensional Gaussian distribution for each.    Additional color charge fluctuations are included on the scale of the lattice spacing within the calculation.   Finally, the deposited energy is calculated.  The \ipglasma code also time evolves the initial color distribution using the Yang-Mills equations of motion, and this moderates the dependence of the 
additional color charge fluctuations on the lattice spacing.  The \ipglasma initial conditions have been successful at matching experimental flow data when used as input to viscous hydrodynamic calculations -- see for example Refs.~\cite{Schenke:2020mbo,Schenke:2010rr}.  

The \ipjazma phenomenological model~\cite{Nagle:2018ybc} was constructed to specifically evaluate initial conditions from the MSTV calculations for small collisions systems which are calculated in the so-called dilute-dense limit of the CGC framework~\cite{Mace:2018yvl,Mace:2018vwq}.   \ipjazma can also calculate initial conditions as the simple product of two-dimensional target and projectile Gaussian distributions, the so-called dense-dense limit.   These calculations provide an almost identical match to the energy deposit initial conditions from the full \ipglasma framework -- see details in Appendix A.  This agreement emphasizes that the key ingredients for the initial geometry are nucleons or sub-nucleons, Gaussian profiles, and taking the local product of these Gaussians.    Features in the \ipglasma model from color domains or ``spiky'' local fluctuations are sub-dominant, and thus not confirmed by agreement with flow data.   We highlight that the \ipglasma model also calculates the early pre-hydrodynamic time evolution, often up to $\tau = 0.4~\mathrm{fm}/c$, and this is not modeled in \ipjazma or \trento.   The evaluation of this pre-hydrodynamic time evolution and its apples-to-apples comparison with free streaming or strongly-coupled dynamics is a topic for another paper.

A new approach was recently put forward also within the CGC framework, termed \magma~\cite{Gelis:2019vzt}.\footnote{In the process of finalizing this manuscript, the \magma authors, Ref.~\cite{Gelis:2019vzt}, pointed out a potential problem with the CGC correlator used in the model.   This issue is under investigation by those authors.}    A recent analysis comparing various models including \magma is given in Ref.~\cite{Floerchinger:2020tjp}.    In the following sections we (a) detail the \magma calculation and reproduce their results, (b) show results from \magma initial conditions run through full hydrodynamic \sonic simulations, (c) show how the \magma results change if hot spots from one nucleus interact with hot spots from the other nucleus -- which is not the default in the \magma framework, and finally (d) detail results from alternative initial condition calculations.

\section{\magma Calculation}

In the \magma framework, each nucleus is modeled as a two-dimensional profile of color charge density calculated within the CGC framework.   The density is built from localized color charges.   What is notable is that these color charges are distributed without any modeling of nucleons, and are only bounded by the total size of the nucleus, characterized by the Woods-Saxon parameters for a Pb nucleus ($R=6.62$~fm and $a=0.55$~fm).  The number of localized color charges used in the \magma calculation is approximately 100 per nuclei, and thus about half the number of nucleons.   The authors highlight that this is distinct from the \ipglasma calculation, where they first distribute the 208 nucleons from the Pb nucleus, and then calculate the saturation momentum depending on the nucleon positions.  It is unclear what physics justification allows for neglecting nucleons for heavy-ion collisions at these energies.   

\begin{figure}
    \centering
    \includegraphics[width=1.0\linewidth]{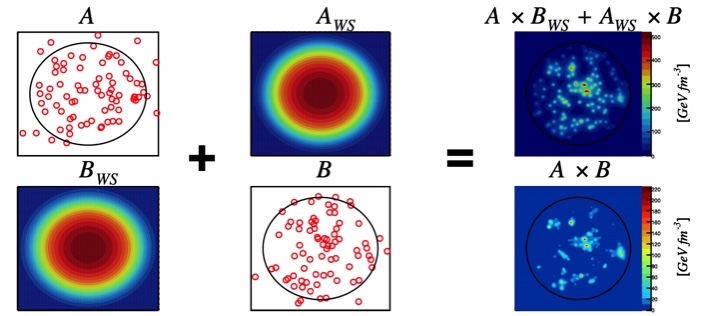}
    \caption{A schematic display of the initial energy deposition calculation.   The hot spots in nucleus A are multiplied by the smooth distribution of nucleus B, labeled as $B_{WS}$ and summed with the inverse, hot spots in nucleus B multiplied by the smooth distribution of nucleus A.   The resulting energy spatial distribution is shown on the top right plot. In contrast, the bottom right plot shows the $A \times B$ calculation using the modified \magma code.
    }
    \label{fig:twoplustwo}
\end{figure}

Another distinguishing feature is that what is calculated in \magma is the energy deposit from localized color charges from the projectile Pb nucleus striking a smooth target nucleus, and then linearly summing the energy deposit from localized color charges from the target Pb nucleus striking a smooth projectile Pb nucleus.    This is nicely visualized in Figure 1 from the \magma paper~\cite{Gelis:2019vzt} -- we have regenerated a version of this representation here as Figure~\ref{fig:twoplustwo}, with the resulting energy deposit shown in the upper right panel.    Each interaction creates a sharply peaked energy deposit that decreases as the distance squared.   The calculation reproduced the one-point and two-point functions of the energy density field calculation in the CGC effective theory~\cite{Albacete:2018bbv}.   We thus label the \magma calculation as ``$A \times B_{WS} + B \times A_{WS}$'', which is in striking contrast from the \ipglasma ``$A \times B$'' calculation for the local energy density with two nuclei $A$ and $B$, with the resulting energy deposit shown in the lower right panel of Figure~\ref{fig:twoplustwo}.    

\begin{figure}[thb]
   \centering
    \includegraphics[width=\linewidth]{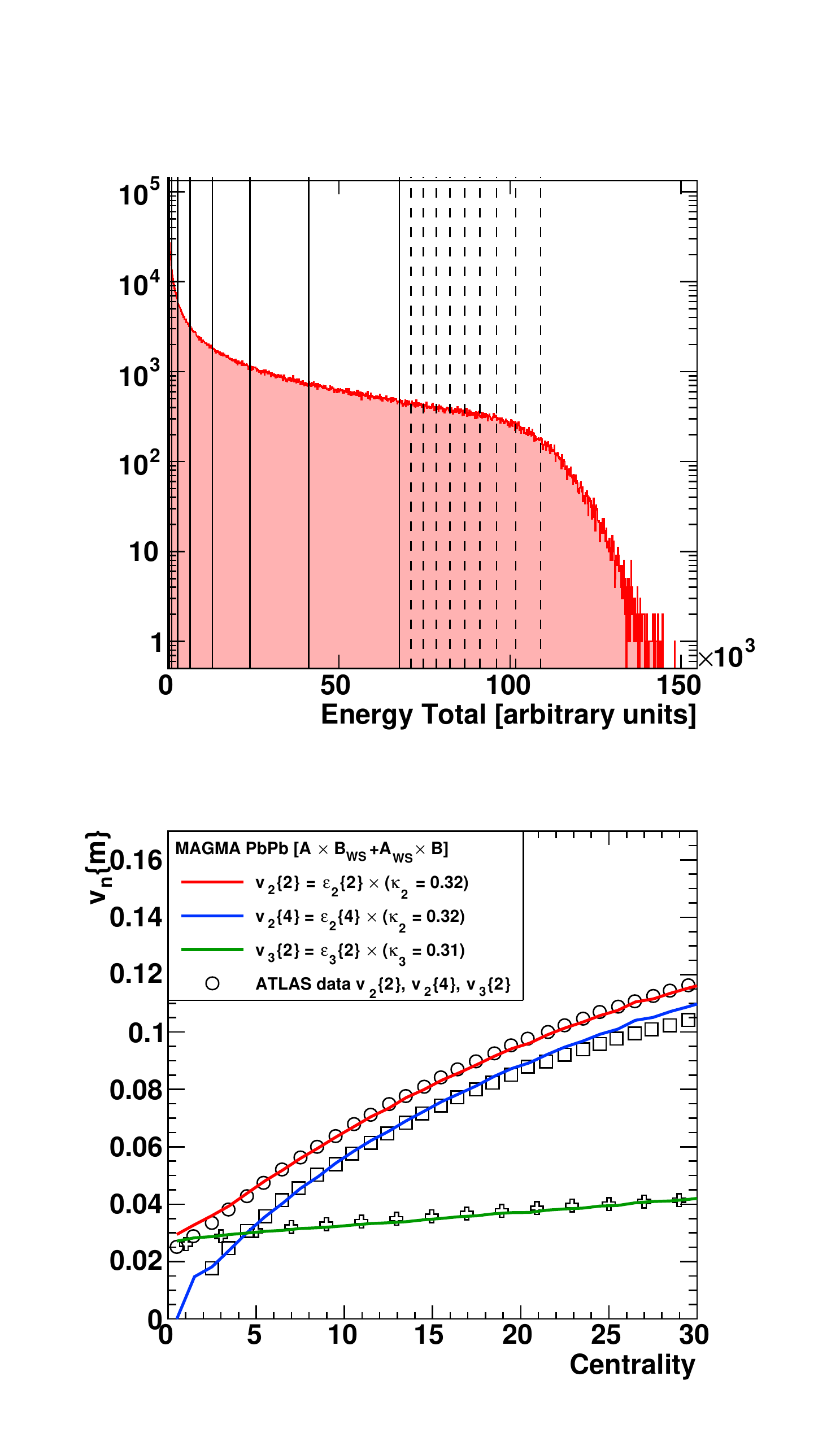}
    \caption{Initial condition calculation with \magma.   (Top) The total energy deposit distribution over one million events.
    Solid (dashed) vertical lines correspond to 10\% (top 1\%) percentiles.
 (Bottom) The $\varepsilon_{2}\{2\}$, $\varepsilon_{2}\{4\}$, and $\varepsilon_{3}\{2\}$ values as a function of centrality selection, scaled up by the respective $\kappa_{2,3}$ values.   In comparison, ATLAS experiment data are shown for $v_{2}\{2\}$, $v_{2}\{4\}$, and $v_{3}\{2\}$.}
    \label{fig:magma}
\end{figure}

Using the publicly available Python code from the \magma authors, we have reproduced their main result, as shown in Figure~\ref{fig:magma}.   The top panel shows the Pb+Pb energy deposit distribution in arbitrary units.    The distribution is dividing into percentiles and then the geometric eccentricities are calculated within the individual centrality selections.    This procedure is not identical to the method of centrality selection in experiment, though we expect this to have negligible impact on our conclusions.  The second- and fourth- cumulants are calculated for the $n=2$ and $n=3$ flow harmonics as follows.

\begin{gather}
\varepsilon_{2}\{2\} = \sqrt{ \left< \varepsilon_{2}^{2} \right> }\\
\varepsilon_{2}\{4\} = (2 \left<\varepsilon_{2}^{2} \right>^{2} - \left<\varepsilon_{2}^{4} \right>)^{1/4}\\
\varepsilon_{3}\{2\} = \sqrt{ \left< \varepsilon_{3}^{2} \right> }
\end{gather}

In order to compare with experimental data, one takes advantage of the fact that hydrodynamics gives an approximately linear relationship between the final flow coefficient and the initial spatial anisotropy (e.g. $v_{2}\{2\} = \kappa_{2} \times \varepsilon_{2}\{2\}$, 
$v_{2}\{4\} = \kappa_{2} \times \varepsilon_{2}\{4\}$, and $v_{3}\{2\} = \kappa_{3} \times \varepsilon_{3}\{2\}$).    We note that this ignores potential contributions from non-linear response~\cite{Betz:2016ayq}, a point we will discuss in the next section.
The $\kappa_{2}$, $\kappa_{3}$ values depend in detail on the QGP properties such as the shear viscosity to entropy density ratio ($\eta$/S) and the treatment of hadronic re-scattering after hydrodynamic expansion.    However, one can assume that these values to vary modestly with collision centrality, and thus they are fitted to experimental data after which one can examine the centrality dependence.    We highlight that the $\kappa$ values are numerically determined by matching the experimental data at centrality = 20, and this is done in a consistent manner in the later comparisons in the paper.   A single value of $\kappa_{2}$ determines the scaling for both the $v_{2}\{2\}$ and $v_{2}\{4\}$.     Values for $\kappa_{2} = 0.32$ and $\kappa_{3}=0.31$ are obtained, in good agreement with the numbers quoted in Ref.~\cite{Gelis:2019vzt}.   

\begin{figure}[thb]
    \centering
    \includegraphics[width=\linewidth]{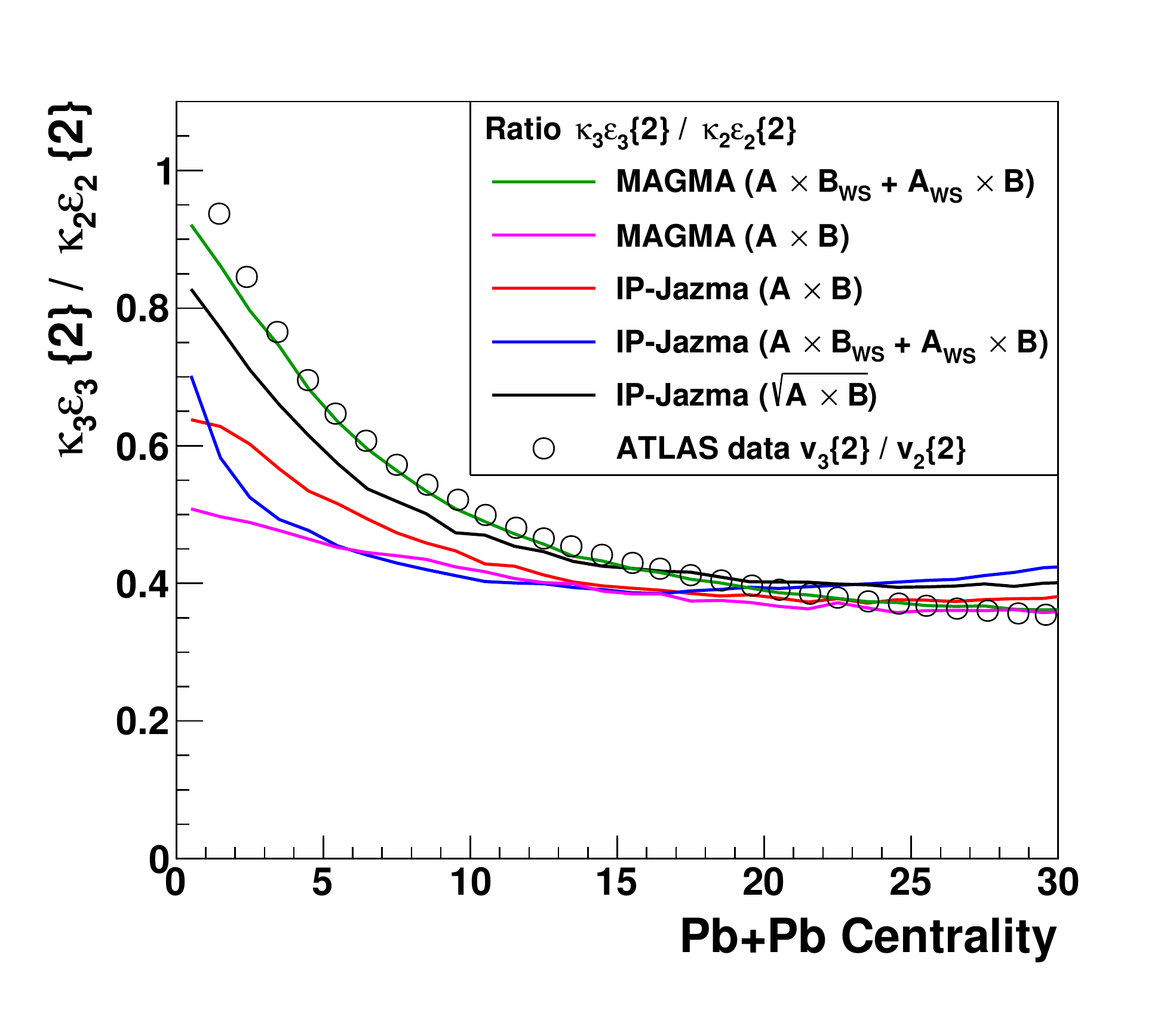}
    \caption{Ratio of $v_{3}\{2\} / v_{2}\{2\}$ as a function of \pbpb collision centrality as measured by the ATLAS experiment.   Multiple theoretical calculations are also shown.}
    \label{fig:ratios}
\end{figure}

Figure~\ref{fig:magma} (above) also shows the experimental data for $v_{2}\{2\}$, $v_{2}\{4\}$, and $v_{3}\{2\}$ as measured by the ATLAS experiment~\cite{Aaboud:2019sma}.
The ATLAS measurements are from charged particles in $0.5<p_{T}<5~\mathrm{GeV}$ and $|\eta|<2.5$ with a requirement of a minimum pseudorapidity gap of 1.67 units.    The agreement between the \magma results and the experimental data is excellent, noting particularly the splitting between $v_{2}\{2\}$ and $v_{2}\{4\}$.   Also remarkable is the agreement with both $v_{2}\{2\}$ and $v_{3}\{2\}$ up to the most central 0-1\% Pb+Pb collisions.    Figure~\ref{fig:ratios} shows the ratio of $\kappa_{3} \times \varepsilon_{3}\{2\}$ / $\kappa_{2} \times \varepsilon_{2}\{2\}$ 
as a function of collision centrality for both data and the \magma calculation.   The ratio in data (and from \magma) approaches unity which encapsulates the ultra-central flow puzzle.    

In the limit of impact parameter $b=0$ Pb+Pb collisions, the average geometry is circularly symmetric and all spatial anisotropies are zero, i.e. $\varepsilon_{n} = 0$.  However, with random fluctuations, for example from nucleon position fluctuations, one obtains non-zero eccentricities but where all moments are approximately equal, i.e. $\varepsilon_{2} \approx \varepsilon_{3} \approx \varepsilon_{4} \approx ... \approx \varepsilon_{n}$~\cite{Mocsy:2011xx}.   However, even in this case, in general the translation of initial geometry into flow is less efficient for higher moments and thus one would expect $\kappa_{3}$ is less than $\kappa_{2}$ -- in contradistinction to the $\kappa_{2,3}$ values obtained in the \magma fit.   For a recent discussion of this puzzle, see Ref.~\cite{Carzon:2020xwp}.   We explore this translation of geometry to flow quantitatively in the next section.

\section{Full hydrodynamic calculations}

The linear factors $\kappa_{2,3}$ can be determined either phenomenologically by matching calculated $\varepsilon_{n}$ to measured $v_{n}$ (as is done in the \magma result shown above) or can be calculated directly with viscous hydrodynamics or parton kinetic theory as examples.
It is striking in the \magma calculation that $\kappa_{2} \approx \kappa_{3}$; i.e. the elliptic and triangular flow have the same linear response coefficient.    In general, even in the case of small viscous damping, e.g. shear viscosity to entropy density $\eta/s = 1/4\pi$, the response coefficient is expected to be smaller for higher moments, i.e. larger $n$ values.    This feature has also been seen in parton transport calculations~\cite{Alver:2010gr}.   We can test this specifically for the 0-1\% \pbpb collision initial conditions from \magma.    

\begin{figure}
    \centering
    \includegraphics[width=1.0\linewidth]{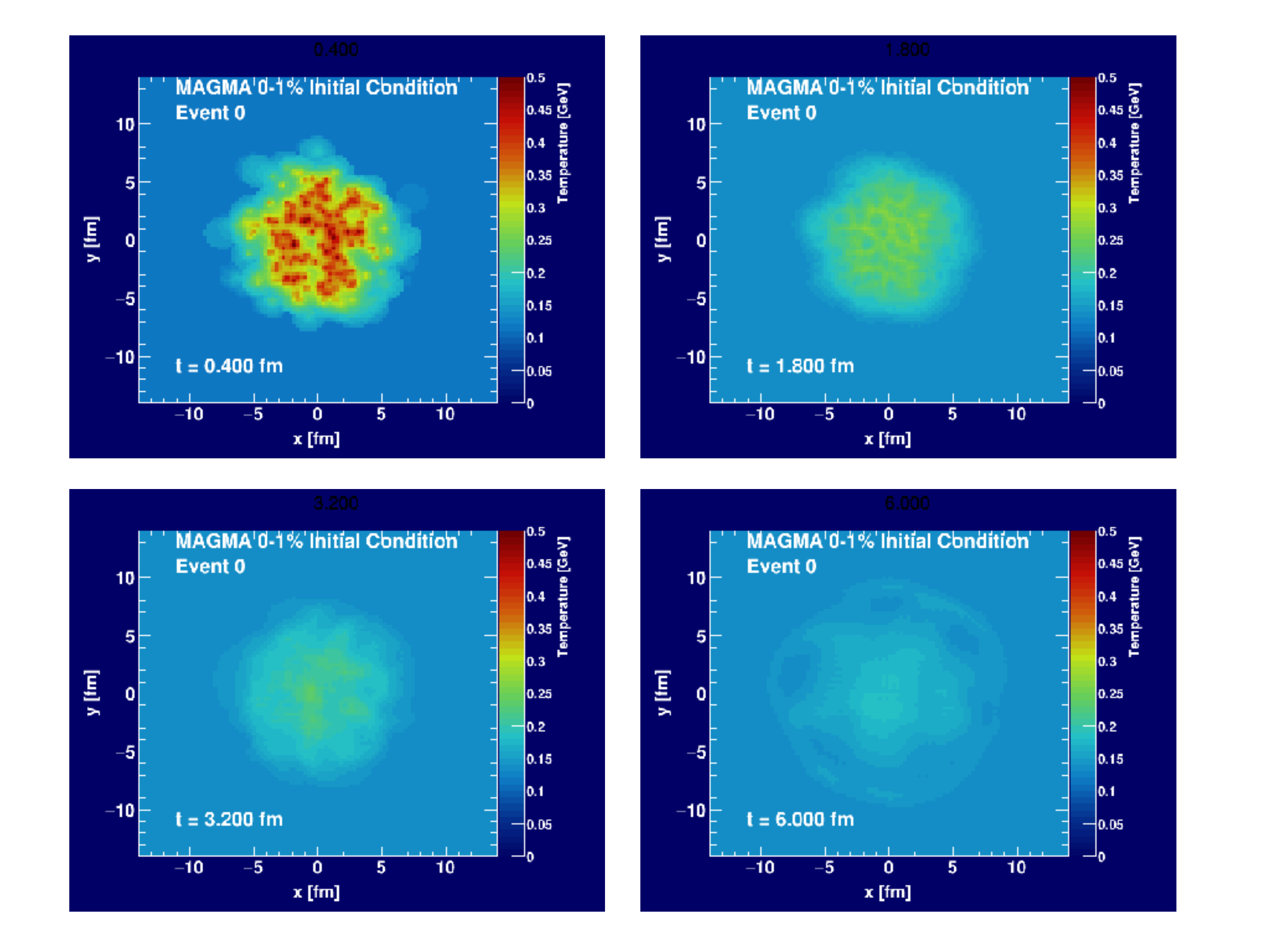}
    \caption{Time evolution event display for a \pbpb 0-1\% most central collision using \magma initial conditions run through \sonic viscous hydrodynamics.  The time snapshots show the temperature in the transverse (x,y) plane and are at $t=0.4, 1.8, 3.2, 6.0$~fm/c.}
    \label{fig:hydrodisplay}
\end{figure}

To this end, we have run 1000 such \magma initial conditions for events all falling into the 0-1\% centrality selection through a full hydrodynamic simulation including hadronic cascade afterburner B3D using the publicly available \sonic code~\cite{Romatschke:2015gxa}.  Figure~\ref{fig:hydrodisplay} shows time snapshots of the two-dimensional temperature profile from a single \magma initial condition through hydrodynamic evolution.   
The \sonic running conditions for the hydrodynamic stage include shear viscosity to entropy density $\eta$/S $= 1/4\pi$ and bulk viscosity to entropy density $\zeta/s = 0$.   The hydrodynamic initial time is set to $\tau_{0} = 0.4$~fm/c and the freeze-out temperature is set to $T_{f} = 170$~MeV.

\begin{figure*}
    \centering
    \includegraphics[width=1.0\linewidth]{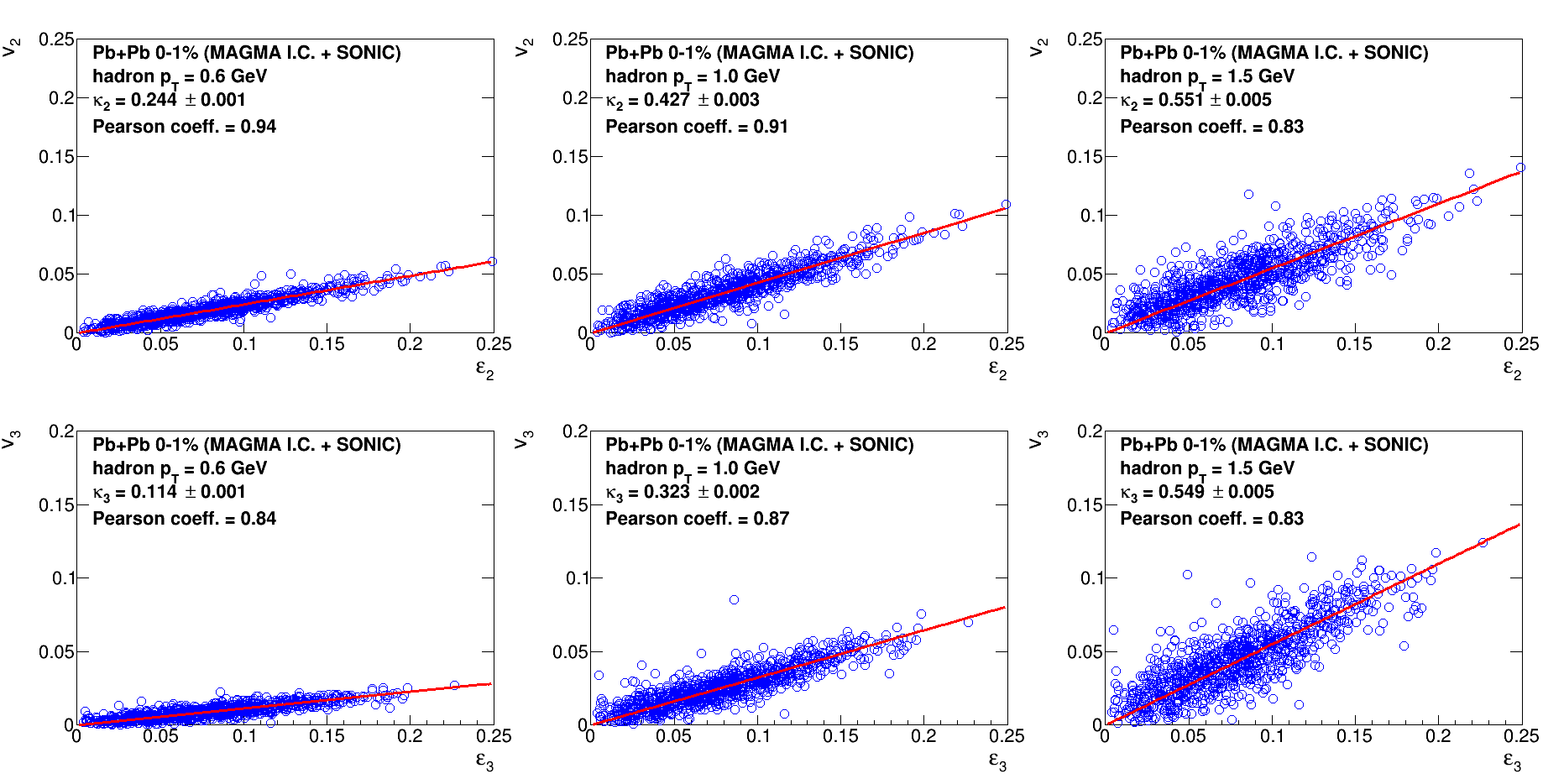}
    \caption{Results from \sonic hydrodynamic calculations for \magma initial condition 0-1\% central Pb+Pb collisions.   Shown are results in the upper (lower) panels from individual events for $v_{2}$ ($v_{3}$) versus the initial geometric $\varepsilon_{2}$ ($\varepsilon_{3}$).   The linear fits represent the $\kappa$ response coefficients.  Also shown are the Pearson coefficients indicating the degree of linear correlation.}
    \label{fig:sonic}
\end{figure*}

\begin{figure*}
    \centering
    \includegraphics[width=1.0\linewidth]{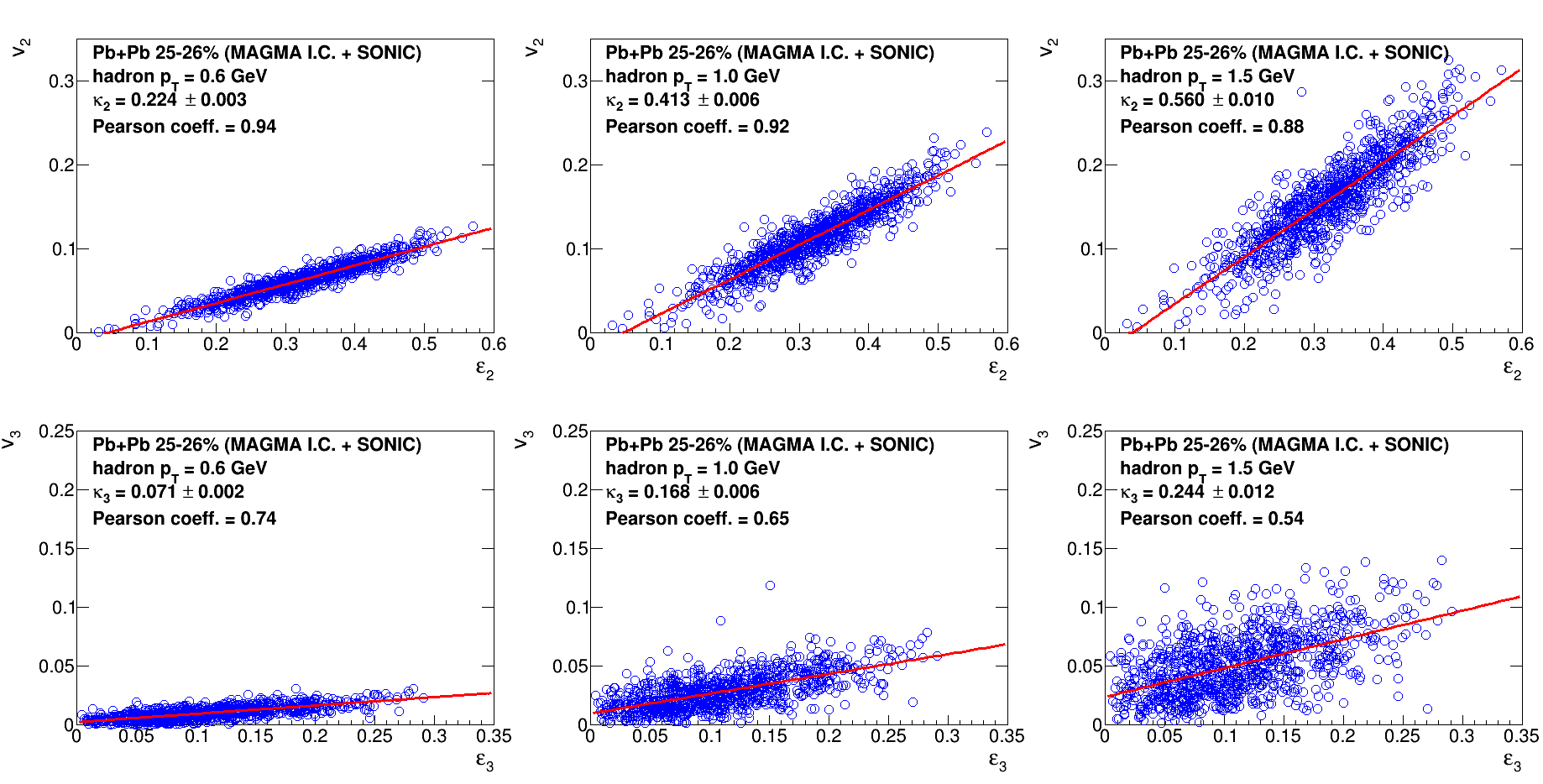}
    \caption{Results from \sonic hydrodynamic calculations for \magma initial condition 25-26\% central Pb+Pb collisions.   Shown are results in the upper (lower) panels from individual events for $v_{2}$ ($v_{3}$) versus the initial geometric $\varepsilon_{2}$ ($\varepsilon_{3}$).   The linear fits represent the $\kappa$ response coefficients.  Also shown are the Pearson coefficients indicating the degree of linear correlation.}
    \label{fig:sonic25}
\end{figure*}

In Figures~\ref{fig:sonic} and~\ref{fig:sonic25}, we plot for \pbpb centralities 0-1\% and 25-26\%, respectively, the flow coefficients $v_{2}$ (upper) and $v_{3}$ (lower) for three different $p_{T}$ selections versus the \magma initial geometry $\varepsilon_{2}$ and $\varepsilon_{3}$ for 1000 individual events.    One sees a reasonable linear relationship in all cases as indicated via the calculated Pearson coefficients shown in the legend.   Each panel is fit to a line with the intercept forced at zero and the slope corresponding to the $\kappa_{n}$ value.      It is noticeable that for the 25-26\% centrality events, where the events extend out to larger values of $\varepsilon_{2}$ there is a clear non-linearity contribution - which is reasonably described by a quadratic fit.    Another observation is that the Pearson coefficients are significantly lower for the $v_{3}$ in the 25-26\% centrality compared with the 0-1\% centrality, i.e. there is a lot more event-to-event spread around the central linear relation.   Lastly, the $\kappa_{n}$ values increase with increasing \pt.  Since the $\varepsilon_{n}$ values for each event do not depend on particle \pt, this increase is simply a reflection of the larger $v_{n}$ as a function of \pt.  
We highlight that in general any linear approximation of flow coefficients with eccentricities is not known to be valid for $p_{T}$-differential anisotropies and thus integrating over a finite $p_{T}$ range introduces a sensitivity on the infrared-cut used~\cite{Luzum:2010ag}.

\begin{figure*}
    \centering
    \includegraphics[width=0.45\linewidth]{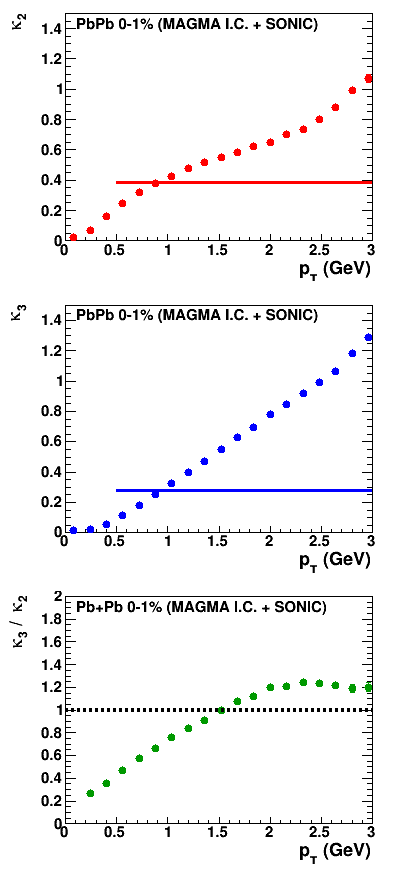}
    \includegraphics[width=0.45\linewidth]{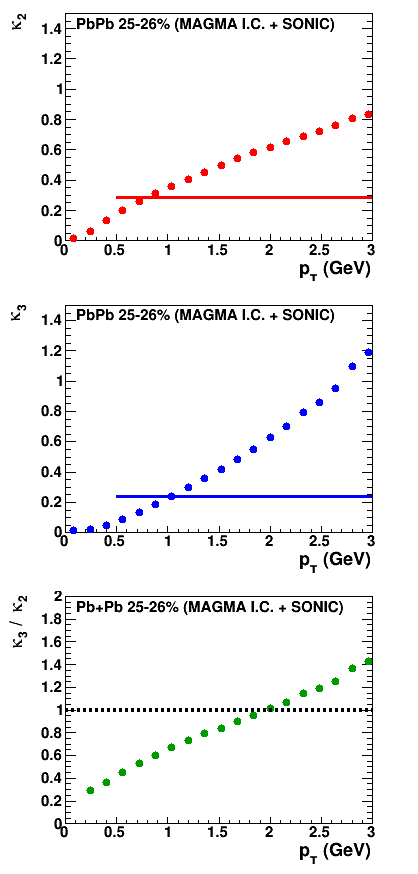}
    \caption{Results from \sonic hydrodynamic calculations for \magma initial condition 0-1\% (left) and 25-26\% (right) Pb+Pb collisions.   Shown are results in the upper (middle) panels are the $\kappa_{2}$ ($\kappa_{3}$) values as a function of charged hadron $p_T$.   The horizontal line represented the $p_{T}$-integrated value over the range 0.5 - 3.0~GeV.   The lower panel shows the ratio of $\kappa_{3}/\kappa_{2}$ as a function of $p_{T}$.  The dashed line is set at one for reference.
}
    \label{fig:sonicpt}
\end{figure*}

Figure~\ref{fig:sonicpt} shows for \pbpb collisions of 0-1\% (left) and 25-26\% (right) centralities the $\kappa_{2}$ (upper) and $\kappa_{3}$ (middle) coefficients and their ratio (lower) as a function of charged hadron $p_{T}$.    The $\kappa_{2}$ and $\kappa_{3}$ $p_{T}$-integrated values over the range 0.5 - 3.0~GeV are shown as solid horizontal lines.   The lower \pt selection is made to match the ATLAS measurement range and the upper \pt selection is nearing the limit where the hydrodynanic calculations has significant systematic uncertainties.    The values for \pbpb 0-1\% centrality are $\kappa_{2} = 0.38$ and $\kappa_{3}=0.28$, and for \pbpb 25-26\% centrality are $\kappa_{2}=0.28$ and $\kappa_{3} = 0.24$.   Thus, the assumption used in the \magma comparison in Figures~\ref{fig:magma} and ~\ref{fig:ratios} of $\kappa_{n}$ independent of centrality is significantly in error.   It is also notable that the ratio of $\kappa_{3}/\kappa_{2}$ varies between these two centrality selections, 0.73 (0-1\%) and 0.85 (25-26\%).   Both of these values are substantially lower than the 0.31/0.32 = 0.97 obtained from the \magma fit shown in Figure~\ref{fig:magma}.

\begin{figure}
    \centering
    \includegraphics[width=0.95\linewidth]{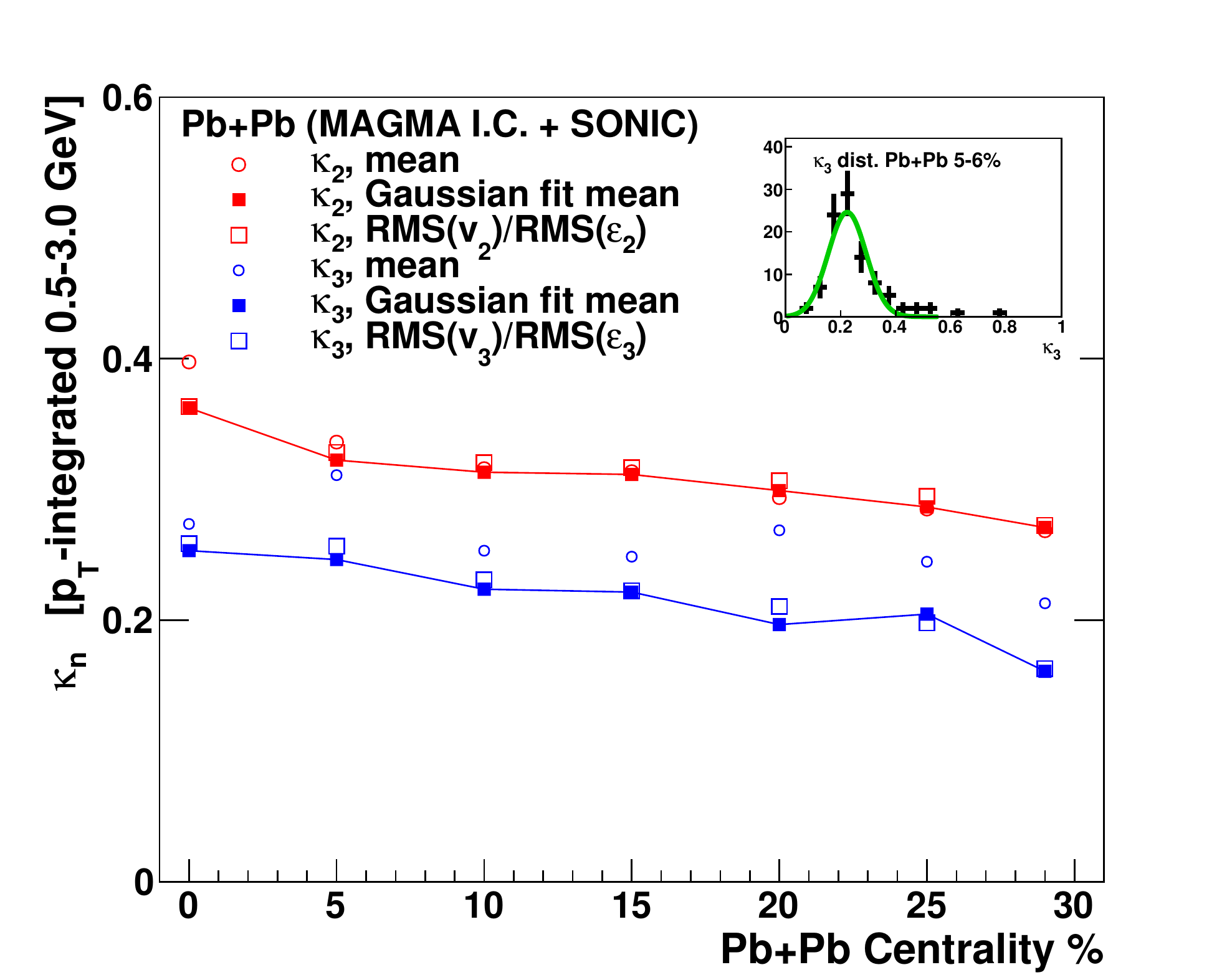}
    \caption{Results from \sonic hydrodynamic calculations for \magma initial condition 0-30\% Pb+Pb collisions quantifying the $\kappa{2}$ and $\kappa_{3}$ values both via the mean values and the Gaussian fit mean values.   Also shown are values of $\kappa_{n}$ determined as the RMS of $v_{n}$ divided by the RMS of $\varepsilon_{n}$.   An example fit is shown in the inset.}
    \label{fig:magmacentdep}
\end{figure}

We have mapped out the $\kappa_{n}$ values over the full centrality range 0-30\% corresponding to charged hadrons with $p_{T} = 0.5-3.0$~GeV as shown in Figure~\ref{fig:magmacentdep}.   The open red (blue) points correspond to the mean $\kappa_{2}$ ($\kappa_{3}$) values.   In the inset, we show the event-by-event distribution of $\kappa_{3}$ values for the specific Pb+Pb 5-6\% centrality.    There is a non-Gaussian high-side tail which is dominated by events with very small values of $\varepsilon_{3}$.   We have also fit these distributions to a Gaussian and shown the Gaussian mean values in Figure~\ref{fig:magmacentdep} as closed points.    There is a clear and substantial centrality dependence for both $\kappa_{2}$ and $\kappa_{3}$ values.   The method for comparison of measured $v_{2}\{2\}$ with $\kappa_{2} \times \varepsilon_{2}\{2\}$ for example, used in the \magma analysis in Figure~\ref{fig:magma}, would be technically more comparable to extracting $\kappa_{2}$ from the \sonic hydrodynamic calculation as the RMS of $v_{n}$ divided by the RMS of $\varepsilon_{n}$.    These values are also shown in Figure~\ref{fig:magmacentdep} and are only very modestly different from the Gaussian mean values.

\begin{figure}
    \centering
    \includegraphics[width=0.95\linewidth]{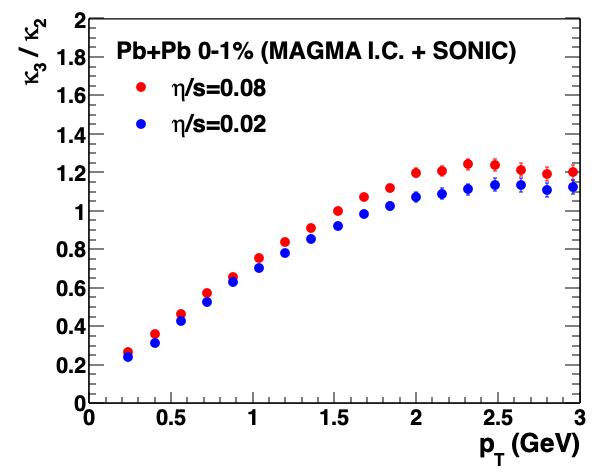}
    \caption{Results from \sonic hydrodynamic calculations for \magma initial conditions 0-1\% for the ratio $\kappa_{3}/\kappa_{2}$ for two different values of $\eta$/S.}
    \label{fig:etaoversratio}
\end{figure}

We have also run calculations with nearly ideal hydrodynamics $\eta/s = 0.02 = 1/4 \times 1/4\pi$ and find for \pbpb 0-1\% centrality value of $\kappa_{2} = 0.53$ and $\kappa_{3}=0.36$.   These are significantly higher than the values quoted above for $\eta/$S = $1/4\pi$ as expected since there is less viscous damping and hence stronger flow.    Shown in Figure~\ref{fig:etaoversratio} is a comparison of the $\kappa_{3}/\kappa_{2}$ ratio with two different values of $\eta/$S.   There are modest difference that again highlight that medium properties do not completely cancel out in these ratios.

The testing of \magma initial conditions with full hydrodynamics reveals that in fact the \magma initial conditions do not match experimental data to resolve the ultra-central puzzle.
Any resolution of the ultra-central puzzle from an initial geometry picture must be coupled with full transport calculations for confirmation.

\section{Alternative \magma Modeling}

\begin{figure}
    \centering
    \includegraphics[width=\linewidth]{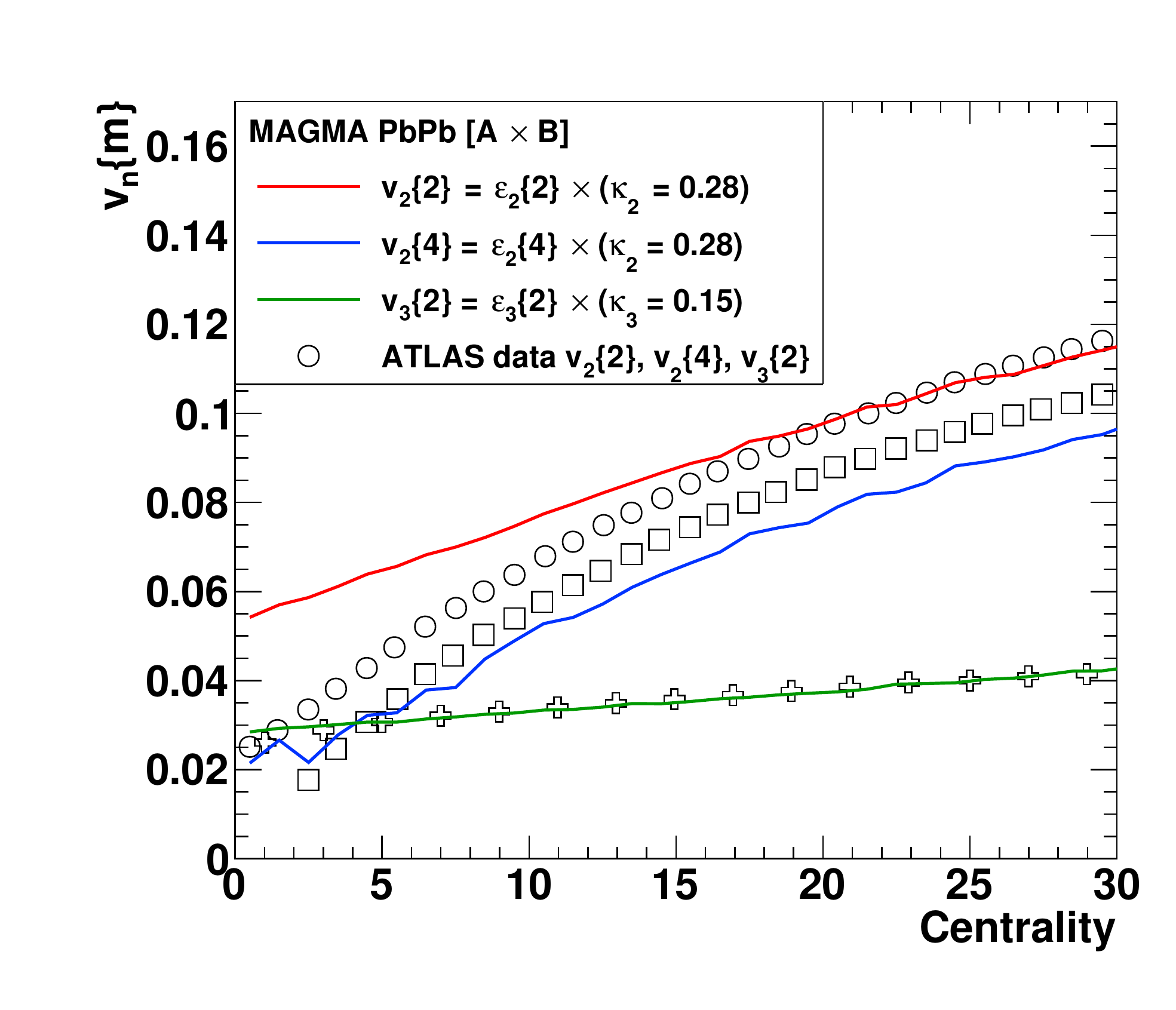}
    \caption{Initial condition calculation with \magma but modified to run where the energy deposit is proportional to $A \times B$. The $\varepsilon_{2}\{2\}$, $\varepsilon_{2}\{4\}$, and $\varepsilon_{3}\{2\}$ values as a function of centrality selection are scaled up by the respective $\kappa_{2,3}$ values.   In comparison, ATLAS experiment data are shown for $v_{2}\{2\}$, $v_{2}\{4\}$, and $v_{3}\{2\}$.}
    \label{fig:magmamod}
\end{figure}

Next we test whether the \magma results are highlight dependent on the non-standard 
$A \times B_{WS} + A_{WS} \times B$ calculation of energy deposit.   To this end, we have modified the \magma code to calculate the energy deposit as $A \times B$, more in line with the weakly-coupled \ipglasma calculation.
Figure~\ref{fig:magmamod} shows the comparison of eccentricity and flow cumulants from the modified-\magma calculation.    The splitting between the $v_{2}\{2\}$ and $v_{2}\{4\}$ is no longer captured by the calculation.  Also the agreement with both $v_{2}\{2\}$ and $v_{3}\{2\}$ is not maintained.   These results plotted as the ratio of 
$v_{3}\{2\}$ / $v_{2}\{2\}$ are also shown in Figure~\ref{fig:ratios} and the modified-\magma calculation only reaches 0.5 in the most central events.    Matching the experimental $v_{n}$ data, the new value for $\kappa_{3} = 0.15$ is now much smaller than $\kappa_{2} = 0.28$.    The original \magma calculation has a larger contribution from the intrinsic geometry, encapsulated in the smooth nuclear distribution, compared to fluctuations.   This modified \magma calculation has relatively larger geometry fluctuations and thus the $v_{2}$ has a flatter centrality dependence and the relative scaling to match $v_{2}$ and $v_{3}$ are very different.   Thus, the \magma results are very sensitive to this non-standard calculation of energy deposit, and do not match experimental data using the more standard $A \times B$ method.

\section{Alternative Initial Conditions}

\begin{figure}
    \centering
    \includegraphics[width = \linewidth]{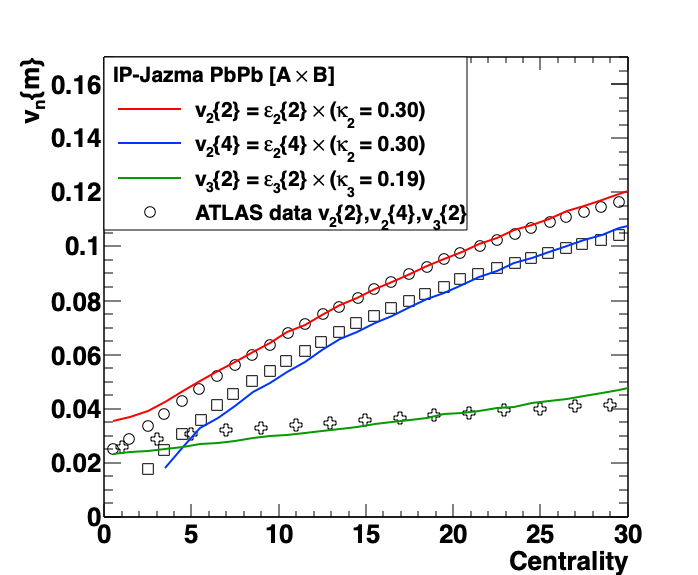}
    \caption{Initial condition calculation with \ipjazma run where the energy deposit is proportional to $A \times B$, as is true in the \ipglasma model.   The $\varepsilon_{2}\{2\}$, $\varepsilon_{2}\{4\}$, and $\varepsilon_{3}\{2\}$ values as a function of centrality selection, scaled up by the respective $\kappa_{2,3}$ values.   In comparison, ATLAS experiment data are shown for $v_{2}\{2\}$, $v_{2}\{4\}$, and $v_{3}\{2\}$.}
    \label{fig:ipjazma0}
\end{figure}

Within the \ipjazma framework, we can calculate initial conditions in a variety of modes.   First, we show results in Figure~\ref{fig:ipjazma0}, where the energy deposit is chosen to be proportional to the local energy density in the projectile times the local energy density in the target ($A \times B$ mode).   As detailed in Appendix A, this models the initial spatial energy distribution in \ipglasma almost perfectly.    The agreement with experimental data is reasonable, although there is more splitting between $v_{2}\{2\}$ and $v_{3}\{2\}$ in the calculation.   Also, as shown as a ratio in Figure~\ref{fig:ratios}, this geometry does not resolve the ultra-central puzzle.   Interestingly, the phenomenologically fitted $\kappa_{2}$ is now 50\% larger than $\kappa_{3}$, more in line with hydrodynamic expectations.   We highlight that this calculation is with nucleons as two-dimensional Gaussians, and ignores sub-nucleons.   Sub-nucleon degrees of freedom in this context have been explored in Ref.~\cite{Loizides:2016djv} and they do not resolve the ultra-central puzzle.

\begin{figure}
    \centering
    \includegraphics[width=\linewidth]{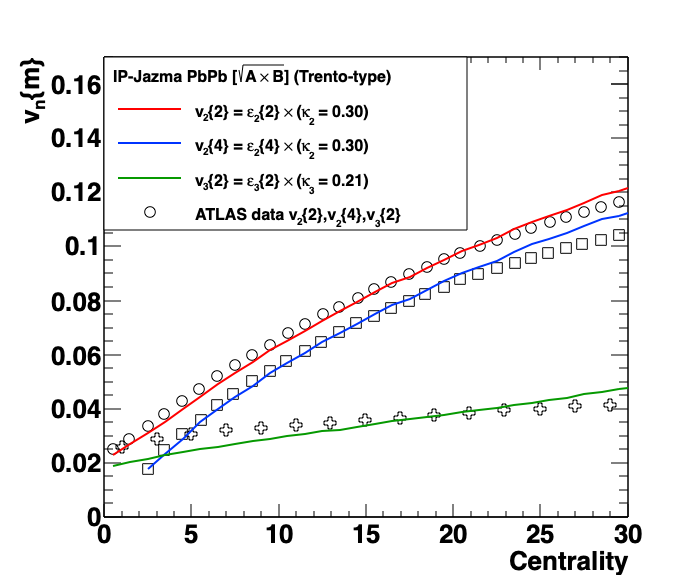}
    \caption{Initial condition calculation with \ipjazma run where the energy deposit is proportional to $\sqrt{A \times B}$ as is true in the \trento model with parameter $p=0$. 
     The $\varepsilon_{2}\{2\}$, $\varepsilon_{2}\{4\}$and $\varepsilon_{3}\{2\}$ values as a function of centrality selection, scaled up by the respective $\kappa_{2,3}$ values.   In comparison, ATLAS experiment data are shown for $v_{2}\{2\}$, $v_{2}\{4\}$ and $v_{3}\{2\}$.}
    \label{fig:ipjazma2}
\end{figure}

Next, we show results in Figure~\ref{fig:ipjazma2}, where the energy deposit is chosen to be proportional to the square root of the local energy density in the projectile times a smooth target nucleus summed with the local energy density in the target times a smooth projectile ($\sqrt{A \times B}$ as done in the \trento calculation with $p=0$).   The agreement with experimental data is quite good, though as shown in Figure~\ref{fig:ratios}, the results are still below the experimental data for $v_{3}\{2\}$ / $v_{2}\{2\}$.    It is again notable that the 
$\kappa_{2}$ is approximately 50\% larger than $\kappa_{3}$, more in line with hydrodynamic expectations.   We note however that these $\kappa_{n}$ values have been extracted by fitting the entire 0-30\% centrality range, and we know from full hydrodynamic simulations that $\kappa_{n}$ vary with centrality.   Thus, a final evaluation can only be made with full hydrodynamic comparison to the data.

In both of these cases, it is important to run full hydrodynamic simulations and with variations on medium properties to have a precision test of the centrality dependence and whether the ultra-central puzzle is reconciled.   Such simulations with \sonic are underway.

\section{Summary}
\label{sec:conclusion}
In summary, we have reproduced the results from the \magma initial condition model and its agreement with elliptic and triangular flow coefficients in \pbpb collisions at the LHC.   However, we find that these results are highly dependent on the energy deposit being proportional to hot spots in the projectile hitting a smooth nuclear target and hot spots in the target hitting a smooth nuclear projectile, i.e. the hot spots do not ``see'' each other.    In addition the translation factors $\kappa_{2,3}$ implied by the data comparison are in contradistinction from what we find with full \sonic hydrodynamic simulations.    A critical take away is that any precision test of initial geometry must be carried out with full evolution to flow coefficients.   We have explored other initial condition modeling (e.g. \ipglasma, \trento) within the \ipjazma framework and find the large triangular flow coefficient in ultra-central \pbpb collisions remains a puzzle requiring further investigation. 

\clearpage

\section{Appendix A}

\begin{figure*}
    \centering
    \includegraphics[width=\linewidth]{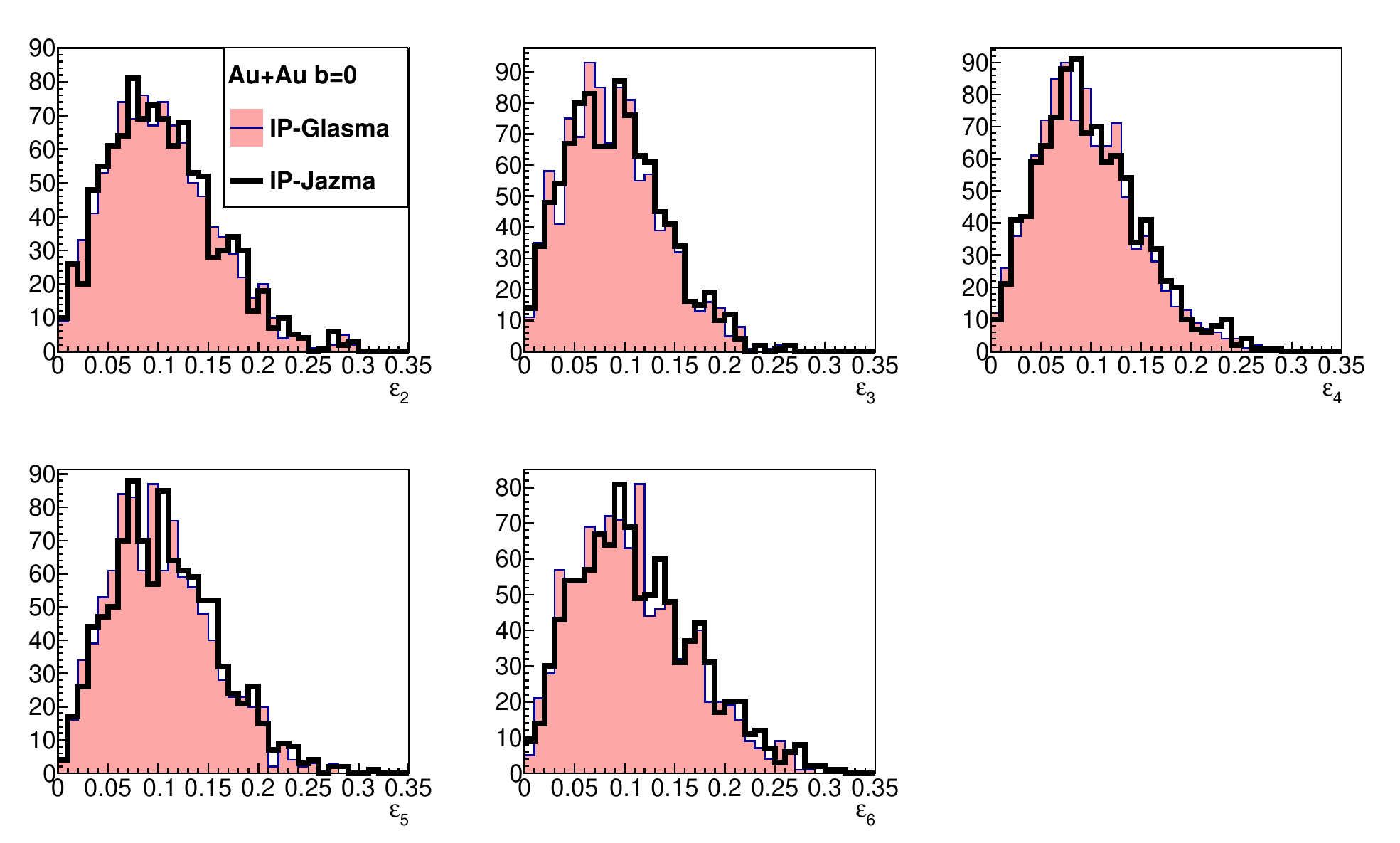}
    \caption{Distributions of spatial eccentricities $\varepsilon_{2-6}$ from \ipglasma and \ipjazma for 1000 Monte Carlo Glauber \auau events of impact parameter $b=0$.
    }
    \label{fig:glasmajazma}
\end{figure*}

As detailed earlier, the \ipglasma is a first principles calculation in the CGC weakly-coupled limit.   Focusing only on the initial distribution of energy deposit in the transverse plane, we find that there are no non-trivial manifestations of color domains or ``spiky'' fluctuations that have been confirmed by experiment through comparisons of flow measurements and initial conditions run through hydrodynamics.   This is in line with studies on the lack of sensitivity to fine-scale structures in long-wavelength hydrodynamics~\cite{Gardim:2017ruc}.
This observation is based on direct comparisons of calculated geometries between two models, \ipjazma and \ipglasma.   As a brief reminder, the \ipjazma calculation has no quantum fluctuations and only has Gaussian distributions associated with nucleons and energy deposit proportional to $A \times B$.   Using identical Monte Carlo Glauber initial conditions for \auau events at $b=0$, fed through the \ipjazma and \ipglasma calculations yield essentially identical $\varepsilon_{2-6}$ distributions -- shown in Figure~\ref{fig:glasmajazma}.   We quantify this comparison for the second and third harmonics with the following values from \ipglasma(\ipjazma):
 $\left< \varepsilon_{2} \right> = 0.104~(0.105)$, $\mathrm{RMS}_{2} = 0.054 ~(0.054)$, Kurtosis$_{2} = 0.271~(0.302)$ and $\left< \varepsilon_{3} \right> = 0.089~(0.091)$, $\mathrm{RMS}_{3} = 0.047~(0.048)$, Kurtosis$_{3} = -0.023~(-0.030)$.
We highlight that in such comparisons it is essential to have the same Monte Carlo Glauber configuration.  For example, the default \ipglasma code has a Glauber two-nucleon exclusion radius $d = 0.8$~fm, which is twice larger than typical values and is unlikely to reflect hard-core repulsion between two nucleons.   If the Monte Carlo Glauber configuration for \ipglasma and \ipjazma were different, one might mistakenly attribute that difference with the models rather than the inputs for the Monte Carlo Glauber.

\begin{figure*}
    \centering
    \includegraphics[width=1.0\linewidth]{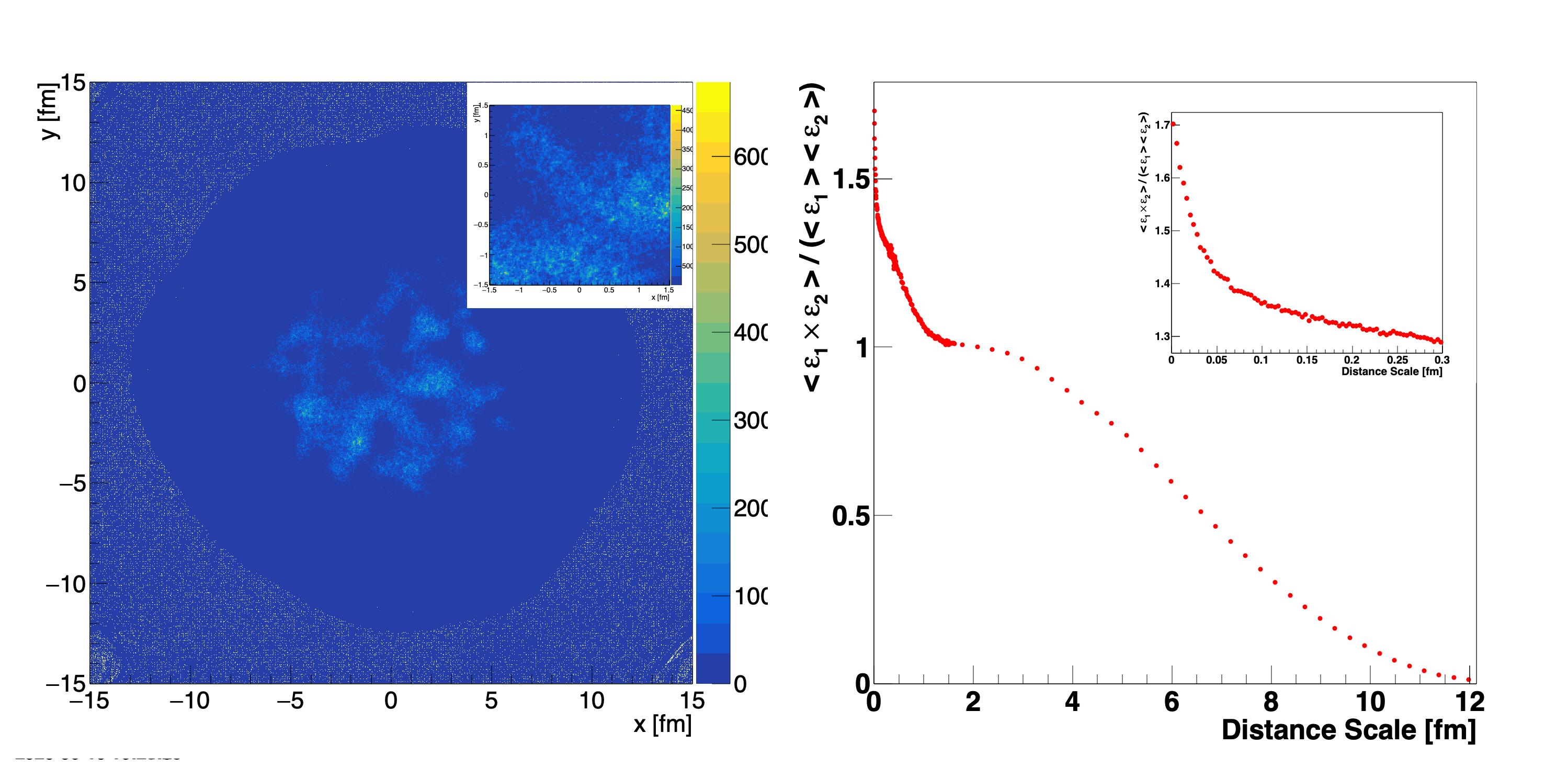}
        \includegraphics[width=1.0\linewidth]{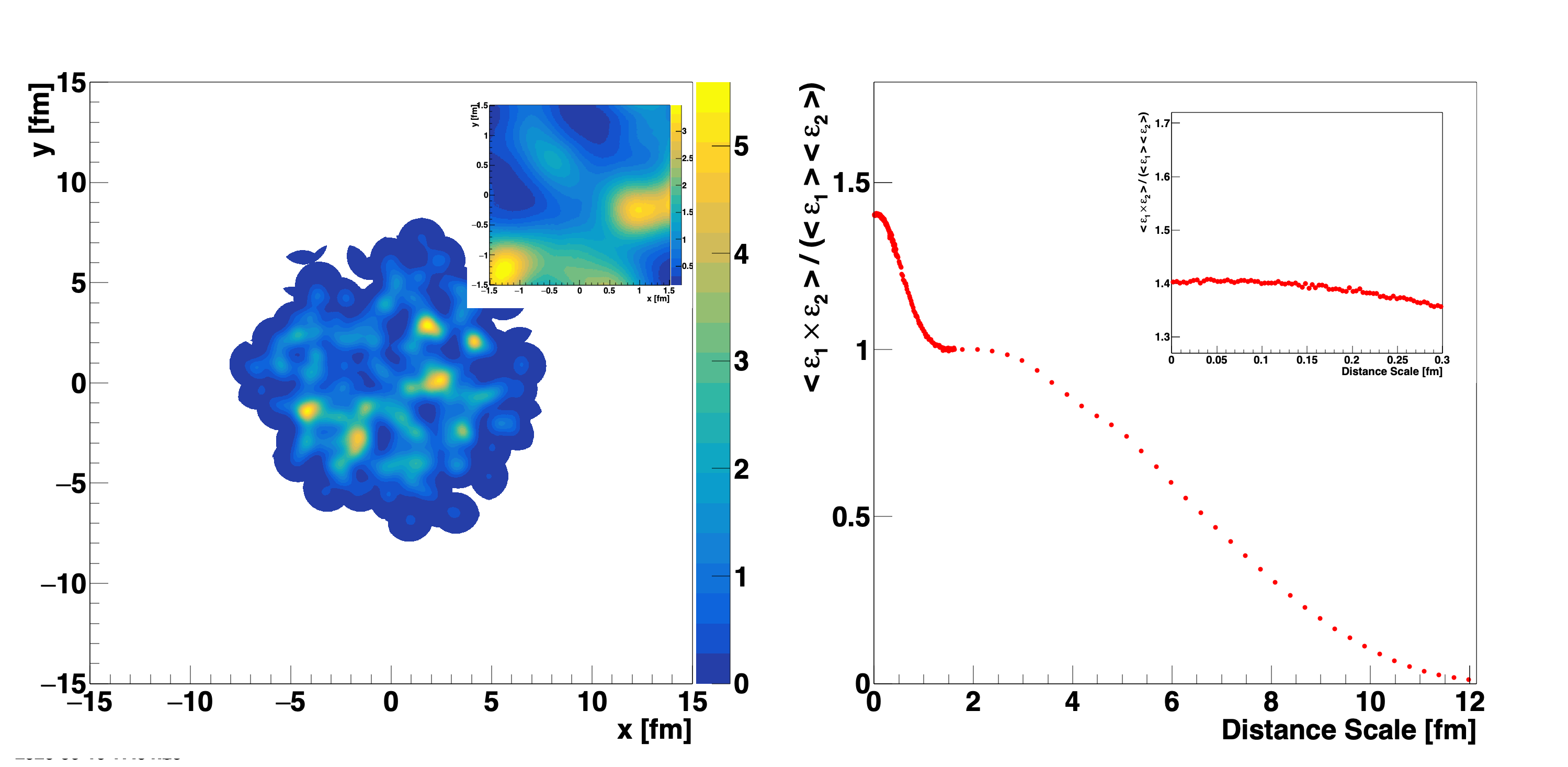}
    \caption{An identical Monte Carlo Glauber \auau event at $b=0$ fed through the \ipglasma (upper panel) and \ipjazma (lower panel) calculations.   The energy density distributions are shown in the left panels, with a zoom inset for more detail.   Also shown are the two-point energy-energy correlator averaged over many events in the \ipglasma (upper panel) and \ipjazma (lower panel) frameworks.}
    \label{fig:glasmajazma2}
\end{figure*}

Another comparison of relevance is the two-point energy-energy correlator.   Shown in the left panels of Figure~\ref{fig:glasmajazma2} are energy deposit displays from \ipglasma (upper panel) and \ipjazma (lower panel) using an identical Monte Carlo Glauber \auau event at $b=0$.   Right panels of Figure~\ref{fig:glasmajazma2} show the correlator ($\left< \varepsilon_{1} \times \varepsilon_{2} \right> / (\left< \varepsilon_{1} \right> \left< \varepsilon_{2} \right>)$), integrated over many events, as a function of distance scale from the \ipglasma (upper panel) and \ipjazma (lower panel) calculations.    One sees a large distance scale (approximately 6~fm) structure from the size of the nucleus and a narrower (approximately 1~fm) structure from nucleons both in \ipglasma and \ipjazma.
Only the narrowest scale structure at $<0.05~\mathrm{fm}$ in the \ipglasma calculation is absent in the \ipjazma calculation -- see the inset for a zoomed in view.   This structure in \ipglasma scales linearly with the lattice spacing used in the calculation and is put in by hand.    Again, it is notable that there is no visible evidence of color domains in the energy deposit structure.   
We note that this ``spiky'' structure gets washed out with the subsequent time evolution in the \ipglasma framework.   \ipjazma is a model only for the initial energy density and does not perform a time evolution as \ipglasma does.

We have run the same comparison with sub-nucleon structure (for example with three constituent quarks) and then there appears another structure in both \ipglasma and \ipjazma (approximately 0.2--0.3~fm), indicating that smaller structures can be seen in principle.   Again, they do not appear to reflect any CGC-specific physics.

\begin{figure*}
    \centering
    \includegraphics[width=\linewidth]{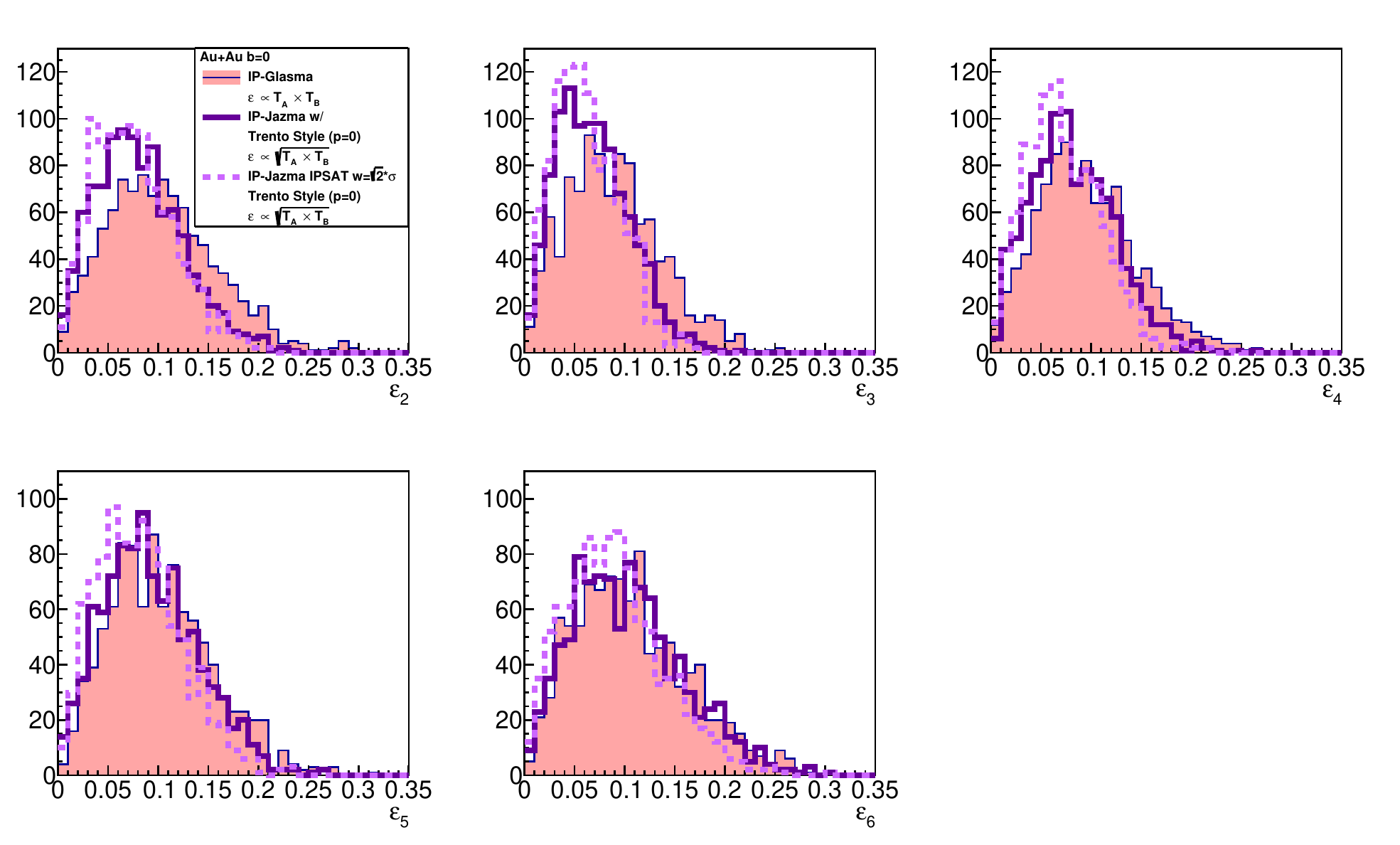}
    \caption{Distributions of spatial eccentricities $\varepsilon_{2-6}$ from \ipglasma and \ipjazma run with the \trento-style energy deposit proportional to $\sqrt{T_{A} \times T_{B}}$ (i.e. $p=0$ \trento mode) for 1000 Monte Carlo Glauber \auau events of impact parameter $b=0$. \ipjazma results where the two-dimensional Gaussian width is adjusted by $\sqrt{2}$ are also shown.}
    \label{fig:glasmajazmatrento}
\end{figure*}

Lastly, we show a comparison of initial geometry from the \ipjazma calculation in a \trento $p=0$ like mode.   The results from \ipjazma with energy deposit proportional to $\sqrt{T_{A} \times T_{B}}$, i.e. the square root of the local nuclear thickness values, are shown in Figure~\ref{fig:glasmajazmatrento}.   Since the energy distribution from each nucleon is distributed as a two-dimensional Gaussian, if one considers that one is taking the square root, i.e.
$\sqrt{T_{A}} \times \sqrt{T_{B}}$, the Gaussian $\sigma$ should be increased by $\sqrt{2}$.   However, one is locally summing the Gaussian contributions from all nucleons in a nucleus and then taking the square root, so there is no perfect match between \trento $p=0$ style and \ipglasma style geometry.

\section*{Acknowledgments}

We gratefully acknowledge useful discussions with Giuliano Giacalone as well as for sharing the \magma Python code.   We also acknowledge useful discussions and a careful reading of the manuscript by Jean-Yves Ollitrault, Paul Romatschke, Anthony Timmins and Bill Zajc.   We acknowledge Bjoern Schenke for the publicly available \ipglasma code and Paul Romatschke for the publicly available \sonic code.   We highlight that the \ipjazma and \trento codes are also publicly available.
RS, MB, JLN acknowledges support from the U.S. Department of Energy, Office of Science, Office of Nuclear Physics under Contract No. DE-FG02-00ER41152.   
SHL acknowledges support from Pusan National University Research Grant, 2019.
\appendix


\clearpage

\bibliography{main}

\end{document}